\documentclass[conference,letterpaper]{IEEEtran}
\IEEEoverridecommandlockouts

\usepackage{cite}

\usepackage{physics}
\usepackage{mathtools,amssymb,lipsum}

\usepackage{cuted}
\usepackage{amsmath,amssymb,amsfonts}
\usepackage{amsthm}
\usepackage{graphicx}
\usepackage{lipsum}
\usepackage{multicol}
\usepackage{relsize}
\usepackage{textcomp}
\usepackage{xcolor}
\usepackage{amsmath, amsthm, amssymb}

\usepackage{url}
\usepackage{caption}
\usepackage{accents}
\usepackage{subfigure}
\usepackage{quantikz}

\newtheorem{lemma}{Lemma}

\usepackage{graphicx}
\usepackage{algorithm}
\usepackage{algpseudocode}
\usepackage{dcolumn}
\usepackage{bm}

\newcommand{\RNum}[1]{\uppercase\expandafter{\romannumeral #1\relax}}
\usepackage{epstopdf}
\usepackage{relsize}

\newcommand{\be}{\begin{enumerate}}
\newcommand{\ee}{\end{enumerate}}
\newcommand{\bd}{\begin{description}}
\newcommand{\ed}{\end{description}}
\newcommand{\bc}{\begin{center}}
\newcommand{\ec}{\end{center}}
\newcommand{\bt}{\begin{tabbing}}
\newcommand{\et}{\end{tabbing}}
\newcommand{\bfig}{\begin{figure}}
\newcommand{\efig}{\end{figure}}
\newcommand{\beq}{\begin{equation}}
\newcommand{\beqarr}{\begin{eqnarray}}
\newcommand{\beqarrn}{\begin{eqnarray*}}
\newcommand{\eeq}{\end{equation}}
\newcommand{\eeqarr}{\end{eqnarray}}
\newcommand{\eeqarrn}{\end{eqnarray*}}
\newcommand{\bflr}{\begin{flushright}\vspace{-0.2in}}
\newcommand{\eflr}{\end{flushright}}
\newcommand{\bsub}{\begin{subequations}}
\newcommand{\esub}{\end{subequations}}

\newcommand{\barr}{\begin{array}}
\newcommand{\earr}{\end{array}}
\newcommand{\nn}{\nonumber}
\newcommand{\ubar}[1]{\underaccent{\bar}{#1}}

\def\BibTeX{{\rm B\kern-.05em{\sc i\kern-.025em b}\kern-.08em
    T\kern-.1667em\lower.7ex\hbox{E}\kern-.125emX}}
\begin{document}

\title{Fault-Tolerant Quantum LDPC Encoders \\
}
 \author{%
  \IEEEauthorblockN{Abhi Kumar Sharma and Shayan  Srinivasa Garani}
  \IEEEauthorblockA{Department of Electronic Systems Engineering, Indian Institute of Science, Bengaluru: 560012, India\\
    Email: \{abhisharma@iisc.ac.in, shayangs@iisc.ac.in\}}
 }



\maketitle

\begin{abstract}
We propose fault-tolerant encoders for quantum low-density parity check (LDPC) codes. By grouping qubits within a quantum code over contiguous blocks and applying preshared entanglement across these blocks, we show how transversal implementation can be realized. The proposed encoder reduces the error propagation while using multi-qubit gates and is applicable for both entanglement-unassisted and entanglement-assisted quantum LDPC codes.  
\end{abstract}
\begin{IEEEkeywords}
	QLDPC codes, fault-tolerant encoders, entanglement-assistance. 
\end{IEEEkeywords}

\section{Introduction}\label{Sec.1}


Low-density parity-check codes introduced by Gallager~\cite{gallager1963low} achieve channel capacities under various considerations. Further, these codes have amenable encoding and decoding algorithms that are now part of very large-scale integration (VLSI) circuits in practice. There are several methods for constructing good families of LDPC codes  \cite{kou2001low,fossorier2004quasicyclic,mackay1999ieee}. Quantum LDPC (QLDPC) based stabilizer \cite{gottesman1997stabilizer} codes are now being investigated for applications in quantum computing and communication systems based on the phenomenal success of their classical counterparts.  

QLDPC codes for correcting independent $X$ and $Z$ errors based on the Calderbank-Shor-Steane (CSS) construction can be constructed using classical codes $C_1$ and $C_2$ such that $C^{\perp}_1\subseteq C_2$. This leads to the presence of short cycles in the Tanner graph of the code, which is detrimental to code performance. However, this problem can be alleviated using entanglement-assisted (EA) \cite{brun2006correcting,nadkarni2021encoding,garani2023theory} codes with error free preshared Einstein Podolksy Rosen (EPR) pairs since the Tanner graph of the un-assisted portion of the QLDPC code is devoid of short cycles. There are many recent works for constructing EA quantum error correcting codes (QECCs) over arbitrary finite fields (see, for instance, \cite{chen2017entanglement,guenda2018constructions, liu2018application,qian2018mds,galindo2019entanglement, luo2016non,nadkarni2021encoding,nadkarni2021entanglement}).      
Hsieh et al. \cite{hsieh2009entanglement} constructed quantum QC-LDPC codes by using a single classical QC-LDPC code. Quantum QC-LDPC codes constructed by Hagiwara et al.~\cite{hagiwara2007quantum} and Pantaleev and Kalachev ~\cite{Panteleev_2022} have short cycles since the code constructions are based on the CSS framework. In \cite{hsieh2009entanglement,Panteleev_2022, hagiwara2007quantum}, the authors did not provide the encoders for the proposed codes. Motivated by the role of entanglement-assistance for improved quantum code designs, we explore such a possibility for designing fault-tolerant encoders, useful to practice.  
 
Fault-tolerant encoding of QLDPC codes is crucial since quantum gate noise is practically significant. The error propagation between the qubits due to multi-qubit gates used in the encoding circuits can lead to erroneous codewords. To alleviate this problem, Gottesman proposed concatenated encoders \cite{gottesman2009introduction} and proved the threshold theorem to reduce error propagation. Later, Hwang \cite{hwang2022faulttolerant} provided algorithms to realize concatenated circuits. However, the concatenated scheme comes at the cost of vanishing code rates with exponentially increasing number of qubits with higher concatenation levels. 

We solve this problem by proposing \textit{transversal} quantum encoders (TQEs) using entangled qubits. The idea is to divide the qubits into blocks, such that quantum operations are performed \textit{locally} in a way that each block is entangled to other blocks using a multipartite entangled state. Each block has a qubit from this preshared entangled state that acts as a control or target for local operations within the block. The overhead in TQEs scales linearly with the number of qubit column blocks and the stabilizer generators of the code, providing better coding rates compared to concatenated codes.  
The paper is organized as follows: In Section \ref{Section:Background}, we describe quantum LDPCs in the CSS framework along with entanglement-assisted codes. In Section \ref{Section:Encoder}, we propose non-fault tolerant encoders and derive a bound on the probability of error propagation. In Section \ref{Section:fault}, we propose fault-tolerant quantum encoders and evaluate probability of error propagation. Finally, we conclude in Section \ref{Sec.5}. 
\section{Background}\label{Section:Background}
In this section, we briefly review the concepts of quantum low-density parity check CSS codes and EA codes. Let $C_1=[n,k_1,d_1]_2$ and $C_2=[n,k_2,d_2]_2$ are the classical codes with parity check (p.c.) matrices $H_1$ and $H_2$ of dimension $\rho_1\times n$ and $\rho_2\times n$ such that the codewords of $C_1$ and $C_2$ lie in the null space of $H_1$ and $H_2$, respectively. The codes $C_1$ and $C_2$ are called LDPCs if the p.c. matrices $H_1$ and $H_2$ are sparse. 
A quantum LDPC code can be generated using $C_1$ and $C_2$ if the codes are dual-containing, i.e., $C^{\perp}_1\subseteq C_2$. For the dual-containing codes, $H_1H^{\mathrm{T}}_2=\textbf{0}\mod 2$. The stabilizer generators of $Q=[[n,k_1+k_2-n,d_q]]_2$ corresponding to the CSS code \cite{gottesman1997stabilizer} are represented as
\begin{equation}
    H_{\mathrm{CSS}}=\left[\begin{array}{c|c}
        H_1 & \textbf{0}_2 \\
        \textbf{0}_1 & H_2
    \end{array}\right],
\end{equation}
where $\textbf{0}_1$ and $\textbf{0}_2$ are zero matrices of dimension $\rho_2\times n$ and $\rho_1\times n$. The code distance $d_q \geq \min\{d_1,d_2\}$. 

The stabilizer group $\mathcal{S}$ can be generated using the generators isomorphic to the rows of the $H_{\mathrm{CSS}}$ matrix such that $I\equiv[0|0]$, $X\equiv[1|0]$, $Z\equiv[0|1]$ and $Y\equiv[1|1]$. If the classical codes are not dual-containing, then we use entanglement-assisted stabilizer codes in which $c=\mathrm{gfrank}(H_1H^{\mathrm{T}}_2)$ number of Bell pairs $\ket{\Phi^{+}}=\frac{1}{\sqrt{2}}(\ket{00}+\ket{11})$ are preshared between the encoder and the decoder. Using these preshared entangled Bell pairs, the dimension of the codespace increases, and the dual-containing criteria is satisfied by obtaining a set of commuting stabilizer generators. The normalizer of a quantum code $Q$, denoted as $\mathcal{N}(Q)$, contains elements that commute with all the elements of the group $\mathcal{S}$. The elements in the $\mathcal{N}(Q)\backslash \mathcal{S}$ is the set of undetected errors that changes the codewords from one to another within the codespace, implying that the minimum distance of the quantum code is the minimum weight of the elements of $\mathcal{N}(Q)\backslash \mathcal{S}$.
\begin{figure}
    \centering
    \begin{equation*}
    \begin{quantikz}[thin lines] 
            \lstick{$\ket{\psi_0}$}&  \gate{X} & \ctrl{1} &  \\
            \lstick{$\ket{0}$} & \qw & \gate{X} &      
    \end{quantikz}
    \Rightarrow
    \begin{quantikz}[thin lines] 
            \lstick{$\ket{\psi_0}$}&   \ctrl{1} & \gate{X} & \\
            \lstick{$\ket{0}$} &  \gate{X} & \gate{X} &     
    \end{quantikz}
\end{equation*}

(a)
  \begin{equation*}
    \begin{quantikz}[thin lines] 
            \lstick{$\ket{\psi_0}$}&  \gate{Z} & \ctrl{1} &  \\
            \lstick{$\ket{0}$} & \qw & \gate{X} &      
    \end{quantikz}
    \Rightarrow
    \begin{quantikz}[thin lines] 
            \lstick{$\ket{\psi_0}$}&   \ctrl{1} & \gate{Z} & \\
            \lstick{$\ket{0}$} &  \gate{X} & \gate{Z} &     
    \end{quantikz}
\end{equation*}

(b)
    \caption{(a) shows the error propagation of an X error to the target qubit when the control qubit is affected with an X error. (b) shows the error propagation of the Z error to the control qubit when the target qubit is affected by the Z error.}
   \label{Fig:1}
\end{figure}
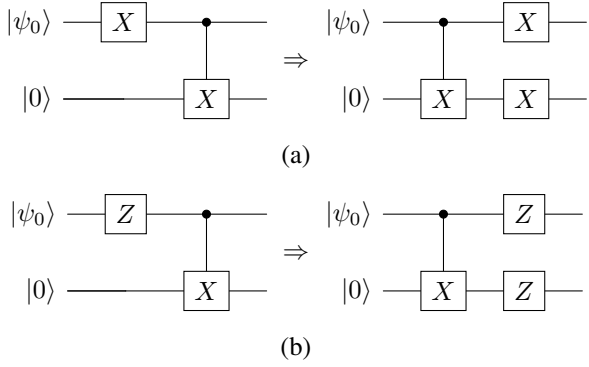

\section{Non-fault-tolerant Encoders}\label{Section:Encoder}

Consider a quantum CSS stabilizer code $Q=[[n,k,d]]_2$ designed using two classical codes $C_1=[n,k_1,d_1]_2$ and $C_2=[n,k_2,d_2]_2$ with parity check matrices $H^{(x)}_{\rho_1\times n}$ and $H^{(z)}_{\rho_2\times n}$ such that $H^{(x)}H^{(z)^\mathrm{T}}=\textbf{0} \mod (2)$. The stabilizer matrix of the quantum code $Q$ is given by
\begin{equation}
    H_{\mathrm{CSS}}=\left[\begin{array}{c|c}
        H^{(x)} & \textbf{0}^{(z)} \\
        \textbf{0}^{(x)} & H^{(z)}
    \end{array}\right],
\end{equation}
where $\textbf{0}^{(z)}$ and $\textbf{0}^{(x)}$ are zero matrices of dimension $\rho_1\times n$ and $\rho_2\times n$.
The non-fault tolerant encoder of quantum CSS code has controlled-NOT (CNOT) gates across all the $n$ qubits. 
To design an encoding operator $\mathcal{E}$, we transform the $H_{\mathrm{CSS}}$ matrix using row operations and columns swapping, i.e., reduce it to row-echelon form as follows:
\begin{equation}
    \Tilde{H}_{\mathrm{CSS}}=\left[\begin{array}{c|c}
        \Tilde{H}^{(x)} & \textbf{0}^{(z)} \\
        \textbf{0}^{(x)} & \Tilde{H}^{(z)}
    \end{array}\right],
    \label{eq:2}
\end{equation}
where $\Tilde{H}^{(x)}=\left[I^{(x)}|A^{(x)}|B^{(x)}\right]$ and $\Tilde{H}^{(z)}=\left[I^{(z)}|A^{(z)}|B^{(z)}\right]$, \\ and $A^{(x)}$ and $B^{(x)}$ $\left(\mathrm{or}\, A^{(z)}\, \mathrm{and}\, B^{(z)}\right)$ are the submatrices of $H^{(x)}$ $\left(\mathrm{or}\, H^{(z)}\right)$ of dimensions $\rho_{1}\times\rho_{2}$ and $\rho_{1}\times(n-\rho_1-\rho_2)$ ($\mathrm{or}\, \rho_{2}\times\rho_{1}$ and  $\,\rho_{2}\times(n-\rho_1-\rho_2)$), respectively.

To obtain the decoding operator $\mathcal{D}$, transform $\Tilde{H}_{\mathrm{CSS}}$ to 

\begin{scriptsize}
\begin{equation}
    \Tilde{H}_{\mathrm{CSS}}=\left[\begin{array}{c|c}
        \begin{array}{cc}
             \textbf{0}^{(z)}_{\rho_1\times\rho_1} & \textbf{0}^{(x)}_{\rho_1\times(n-\rho_1)}\\
              \textbf{0}^{(z)}_{\rho_2\times\rho_1}&\textbf{0}^{(x)}_{\rho_2\times(n-\rho_1)}
        \end{array} &  \begin{array}{ccc}
            I_{\rho_1} & \textbf{0}^{(z)}_{\rho_1 \times \rho_2} & \textbf{0}^{(z)}_{\rho_1 \times w}\\
             \textbf{0}^{(x)}_{\rho_2 \times \rho_1}  & I_{\rho_2} & \textbf{0}^{(z)}_{\rho_2 \times w}
        \end{array} 
    \end{array}\right],
    \label{eq:3}
\end{equation}
\end{scriptsize}
where $w=n-\rho_1-\rho_2$. Due to reversibility, we can obtain the required encoding operator as $\mathcal{E}=\mathcal{D}^{\dagger}$.

From equation (\ref{eq:3}) after applying $\mathcal{D}$ all $X$ stabilizers are identity. Z stabilizers are applied only on the first $\rho_1+\rho_2$ qubits where we have ancilla qubits in state $\ket{0}$ stabilized by Z stabilizers. The last $n-\rho_1 -\rho_2$ qubits are information qubits which are initially unknown and stabilized by I operators.
\subsection{Transformation of X stabilizers}
To make X stabilizers zero in $\Tilde{H}_{\mathrm{CSS}}$, we iterate over the rows of $\Tilde{H}^{(x)}$ to make submatrices $\left[A^{(x)}|B^{(x)}\right]$ zero. For the $i^{\mathrm{th}}$ iteration, we apply $U_i=\mathrm{CNOT}^{(i,C_i)}$ gate, where $i$ is the position of the control qubit and $C_i$ are the nonzero columns of $\left[A^{(x)}_{i,*}|B^{(x)}_{i,*}\right]$, indicating the positions of the target qubits for the CNOTs in $U_i$. Since Pauli X on the control qubit will also be transferred to all the target qubits, in matrix form, $I^{(x)}_{*, i}$~\footnote{Any operator $U$ applied on an $n$-qubit code will change all the stabilizer generators of the code as follows $US_i U^{\dagger}$ for all $i$, so the operator $U$ has performed a column operation in the stabilizer matrix. For example, if $U=\mathrm{CNOT}^{(i,j)}$, then $H^{(x)}_{*,j}$ is replaced with $H^{(x)}_{*,j}+H^{(x)}_{*,i}$ and $H^{(z)}_{*,i}$ with $H^{(z)}_{*,i}+H^{(z)}_{*,j}$.} column will be added to all the columns indicated in $C_i$ to make the $\left[A^{(x)}_{i,*}|B^{(x)}_{i,*}\right]$ zero over $\mathbb{F}_2$ since $I^{(x)}_{i,i}=1$. 

The operator $U=\prod_{i=1}^{\rho_1} U_i$ will make $\left[A^{(x)}|B^{(x)}\right]$ submatrix zero. In addition, the operator $U$ will also affect the Z stabilizers and make first $\rho_1$ columns of $H^{(z)}$ zero. This is proved in the Appendix \ref{Appendix:1}. Now, we apply Hadamard gates on the first $\rho_1$ qubits, i.e.,   $T=\prod_{i=1}^{\rho_1}\mathrm{H}^{(i)}$. As $\mathrm{H}^{(i)}$ gate applied on $i^{\mathrm{th}}$ qubit will swap the $\left[\Tilde{H}^{(x)^{\mathrm{T}}}_{*,i}|\textbf{0}^{(x)^{\mathrm{T}}}_{*,i}\right]^{\mathrm{T}}$ with $\left[\textbf{0}^{(z)^{\mathrm{T}}}_{*,i}|\Tilde{H}^{(z)^{\mathrm{T}}}_{*,i}\right]^{\mathrm{T}}$, from equation (\ref{eq:2}), the $T$ gate will make the submatrix $I^{(x)}$ zero by swapping with the first $\rho_1$ columns of $\textbf{0}^{(z)}$.
\subsection{Transformation of Z stabilizers}
Finally, we solve a linear equation $A^{(z)}P=B^{(z)}$ to obtain the matrix $P$ to make $B^{(z)}$ submatrix zero. Using $P$, we apply $W=\prod_{i=w_1+1}^{n} \mathrm{CNOT}^{(i,O_i)}$, where $w_1=\rho_1 +\rho_2$ and $O_i=\{j:P_{j,i}=1\}$. The operator $W$ will make $B^{(z)}$ matrix zero since the
addition of the columns in $O_i$ of the $A^{(z)}$ submatrix will create the column $\ubar{s}_i=B^{(z)}_{*,i}$ and this $\ubar{s}_i$ will be added to the control column $B^{(z)}_{*,i}$ of $B^{(z)}$ matrix and make it zero. The complete decoding operator of the quantum CSS code is $\mathcal{D}=W T U$ and the encoding operator is $\mathcal{E}=\mathcal{D}^{\dagger}$.\\
\begin{figure*}
    \centering
    \includegraphics[width=0.86\textwidth]{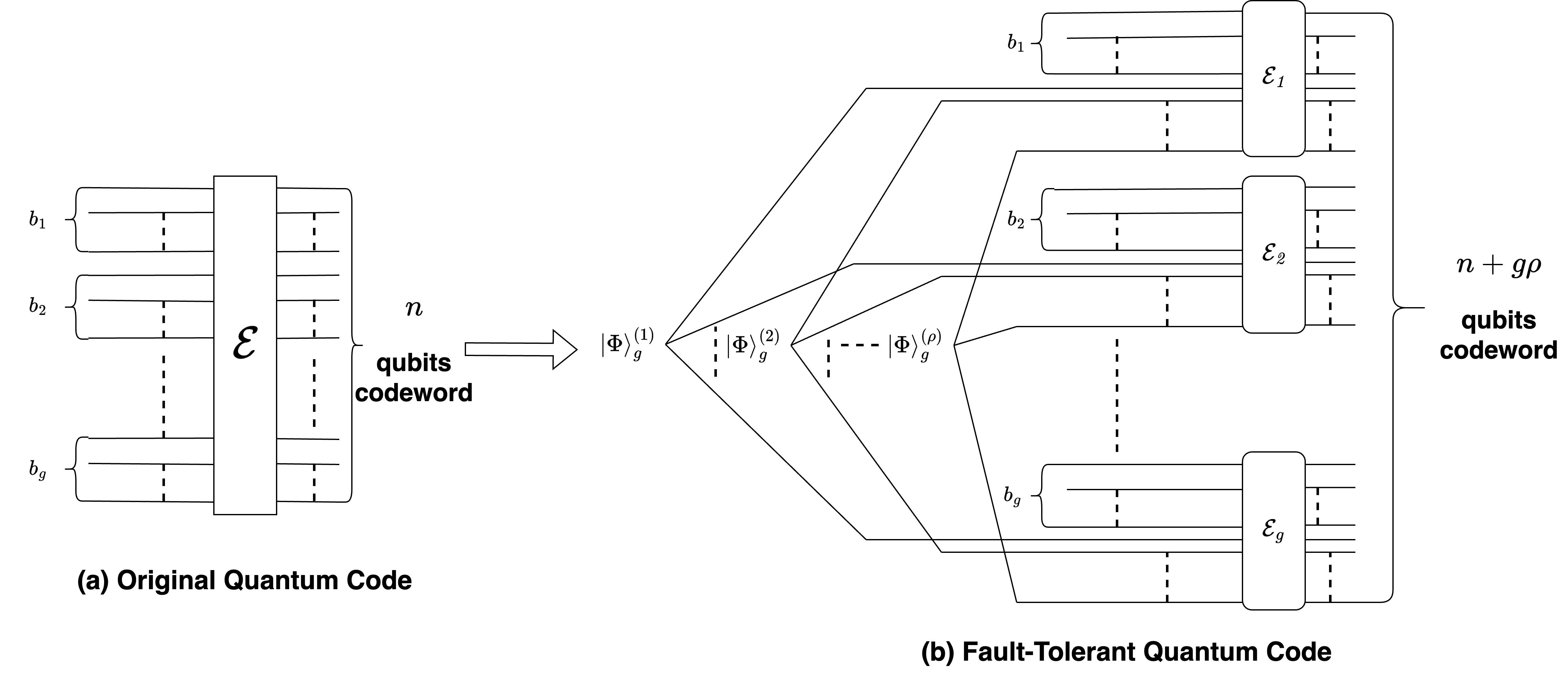}
    \caption{(a) shows the encoding of a non-fault-tolerant quantum code $Q=[[n,k,d]]_2$ such that the operator $\mathcal{E}$ are applied on all $n$-qubits to generate a codeword. (b) shows the fault-tolerant encoding of an modified quantum code $Q_F=[[n+g\rho,k+g\rho,d]]_2$ such that the qubits are divided into the blocks $B_i$ for all $i\in\{1,2,\cdots,g\}$. For every block $B_i$, an $i^{\mathrm{th}}$ entangled qubit of each $\ket{\Phi}^{(i)}_g$ state is associated to design a local encoder $\mathcal{E}_i$ and produces a $n+g\rho$ qubits codeword such that any encoder $\mathcal{E}_i$ does not connect with the $\mathcal{E}_j$ for $i\neq j$ through a multi-qubit gate, so there is no error propagation between any local encoder $\mathcal{E}_i$.}
    \label{fig:2}
\end{figure*}
\subsection{Error Propagation Analysis \\}
The encoder's performance can be measured by calculating the probability of error propagation due to multi-qubit gates.
The popular model of error propagation in case of CNOT gates is shown in Figure \ref{Fig:1}. The X error on the control qubit will be propagated to the target qubit, and the Z error on the target qubit will be propagated to the control qubit.

We consider the depolarizing  noise channel, denoted as $\pi(p)$, which will apply $X$, $Y$ and $Z$ Pauli errors with probability $\frac{p}{3}$ to assess the error propagation performance of the non-fault-tolerant encoder. The probability of error propagation is derived in the following lemma.
\begin{lemma} \label{lemma:5} For a $[[n,k,d]]_2$ CSS code with X and Z stabilizer matrices $\Tilde{H}^{(x)}_{\rho_1 \times n}$ and $\Tilde{H}^{(z)}_{\rho_2 \times n}$, the probability of error propagation using CNOT gates for a non-fault tolerant encoder is upper bounded as $P_{NF} \leq 1 -P_{NF_1} P_{NF_2}$, where
\begin{scriptsize}
\begin{align*}
    P_{NF_1} &\geq \prod_{i=1}^{\rho_1} \left(1-\frac{2p}{3} + \sum_{c \in M^{(AB)}_i} {w^{(AB)}_i \choose c} \left(\frac{2p}{3}\right)^c \left(1-\frac{2p}{3}\right)^{w^{(AB)}_i-c}\right),\\
    P_{NF_2} &\geq \prod_{i=1}^{w} \left(1-\frac{2p}{3} + \sum_{c \in M^{(A)}_i} {w^{(A)}_i \choose c} \left(\frac{2p}{3}\right)^c \left(1-\frac{2p}{3}\right)^{w^{(A)}_i-c}\right),
\end{align*}
\end{scriptsize}
$p$ is the probability of depolarizing noise channel $\pi(p)$, $w^{(AB)}_i$ $\left(\mathrm{or}\, w^{(A)}_i\right)$ is the Hamming weight of $\left[A^{(x)}_{i,*}|B^{(x)}_{i,*}\right]$ $\left(\mathrm{or}\, A^{(z)}_{i,*}\right)$ and $M^{(AB)}_i=\{0,2,\cdots ,w^{(AB)}_i\}\, \left( \mathrm{or}\,M^{(A)}_i=\{0,2,\cdots ,w^{(A)}_i\}\right)$.

\end{lemma}
\begin{IEEEproof}
We prove Lemma \ref{lemma:5} lemma in the Appendix of arxiv version of the paper.
\end{IEEEproof}
The main issue with the non-fault-tolerant encoder introduced in Section \ref{Section:Encoder} is the overlapping application of CNOT gates, which implies that if many CNOT gates have the same target qubit, then a Z or Y error on the target qubit causes propagation of error to all the control qubits. Similarly, if the control qubit of any CNOT gate has an X or Y Pauli error, then the error will be propagated to all the target qubits depending on the Hamming weight $w^{(AB)}_i$ of $[A^{(x)}_i|B^{(x)}_i]$. 

This problem can be solved by reducing $w^{(AB)}_i$ and overlapping CNOTs in an encoder, motivating the design of the fault-tolerant encoder that we describe next.

\section{Fault-tolerant encoders}\label{Section:fault}

 The central ideas behind the transversal realization of the encoder towards fault-tolerance is as follows: (a) First, we divide qubits in the quantum codeword state $\ket{\psi}$ into $g$ contiguous blocks so that controlled gates are applied \textit{locally} within each block using a qubit of the fully entangled multipartite state $\ket{\Phi}_g=\frac{1}{\sqrt{2}}\left(\ket{0}^{\otimes g}+\ket{1}^{\otimes g}\right)$. (b) CNOT gates are \textit{restricted} to each block locally such that a shared qubit of $\ket{\Phi}_g$, assumed to be error free, acts as the control or the target qubit. From the extended part of the entanglement-assisted parity check matrix, further reductions are done using Clifford operations to derive the encoder. The reader must note that error propagation is mitigated since faults do not propagate across blocks. We need a total of $\rho=\max\{\rho_1,\rho_2\}$  $\ket{\Phi}_g$  entangled pairs for our fault-tolerant encoder. We now describe the details behind the design of the fault-tolerant encoder.  
 
 We will demonstrate through an example the idea of fault-tolerant computing.
 Consider two classical LDPC codes $C_1=[9,6,2]_2$ and $C_2=[9,6,2]_2$ with parity check matrices $H_1$ and $H_2$ such that $H_1H_2^{\mathrm{T}}\neq \textbf{0}\mod 2$, where 
\begin{scriptsize}
    \begin{align}
     H^{(x)}=&\left[\begin{array}{ccc|ccc|ccc}
          1 & 0 & 0  & 0 & 1 & 0 & 0 & 0 & 1   \\
        0 & 1 & 0  & 0 & 0 & 1 & 1 & 0 & 0\\
        0 & 0 & 1  & 1 & 0 & 0 & 0 & 1 & 0\\
        \end{array}\right],\nn\\
      H^{(z)}=&\left[\begin{array}{ccc|ccc|ccc}
        1 & 0 & 0 & 0 & 0 & 1  & 0 & 1 & 0   \\
        0 & 1 & 0 & 1 & 0 & 0  & 0 & 0 & 1  \\
        0 & 0 & 1  & 0 & 1 & 0  & 1 & 0 & 0 
      \end{array}\right].
\end{align}
\end{scriptsize}
        
       

Since the $\mathrm{gfrank}(H_1H^{\mathrm{T}}_2)$ is 1, so one Bell pair is required to satisfy the CSS condition, as described in Section \ref{Section:Background}. 
Using $C_1$ and $C_2$, we can design an entanglement-assisted stabilizer code $Q=[[9,4,2;1]]_2$ such that the X and Z stabilizer matrices are designed to satisfy the CSS condition as follows:
\begin{scriptsize}
    \begin{align}
     H^{(x)}=&\left[\begin{array}{ccc|ccc|ccc|c}
          1 & 0 & 0  & 0 & 1 & 0 & 0 & 0 & 1 &1  \\
        0 & 1 & 0  & 0 & 0 & 1 & 1 & 0 & 0&1\\
        0 & 0 & 1  & 1 & 0 & 0 & 0 & 1 & 0&1\\
        \end{array}\right],\nn\\
      H^{(z)}=&\left[\begin{array}{ccc|ccc|ccc|c}
        1 & 0 & 0 & 0 & 0 & 1  & 0 & 1 & 0 &1  \\
        0 & 1 & 0 & 1 & 0 & 0  & 0 & 0 & 1  &1\\
        0 & 0 & 1  & 0 & 1 & 0  & 1 & 0 & 0 &1
      \end{array}\right].
\end{align}
\end{scriptsize}

To design a fault-tolerant quantum encoder, we decide the value of $g$ depending upon the bounds given in equations (\ref{eq:19}). 
For simplicity, we decide to divide the qubits into two blocks, i.e., $g=2$
such that the first block $B_1=\{1,2,\cdots,6\}$ and second block $B_2=\{7,8,9\}$. 

Before encoding, the initial codeword state $\ket{\Psi}$ contains $\{1,3,5\}$ ancillae in $\ket{+}$ state, $\{2,4,6\}$ ancillae in $\ket{0}$ state, $\{7,8\}$ is the information qubits and the last $\{9,10\}$ is an epair shared between transmitter and receiver. The X and Z stabilizer matrices for the state $\ket{\Psi}$ are
\begin{scriptsize}
    \begin{align}
     H^{(x)}&=\left[\begin{array}{cccccc||ccc|||c}
       1 & 0 & 0  & 0 & 0 & 0 & 0 & 0 & 1 &1 \\
        0 & 0 & 1  & 0 & 0 & 0 & 0 & 0 & 1&1\\
        0 & 0 & 0  & 0 & 1 & 0 & 0 & 0&1&1\\
        \end{array}\right],\nn\\
      H^{(z)}&=\left[\begin{array}{cccccc||ccc|||c}
       0 & 1 & 0  & 0 & 0 & 0 &  0 & 0 & 1 &1 \\
        0 & 0 & 0  & 1 & 0 & 0 & 0 & 0 & 1&1\\
        0 & 0 & 0  & 0 & 0 & 1  & 0 & 0&1&1\\
        \end{array}\right],
        \label{Eq:Befor_enc_mat}
        \end{align}
\end{scriptsize}
\begin{figure*}[t]
    \centering    \includegraphics[width=0.6\textwidth, height=3in]{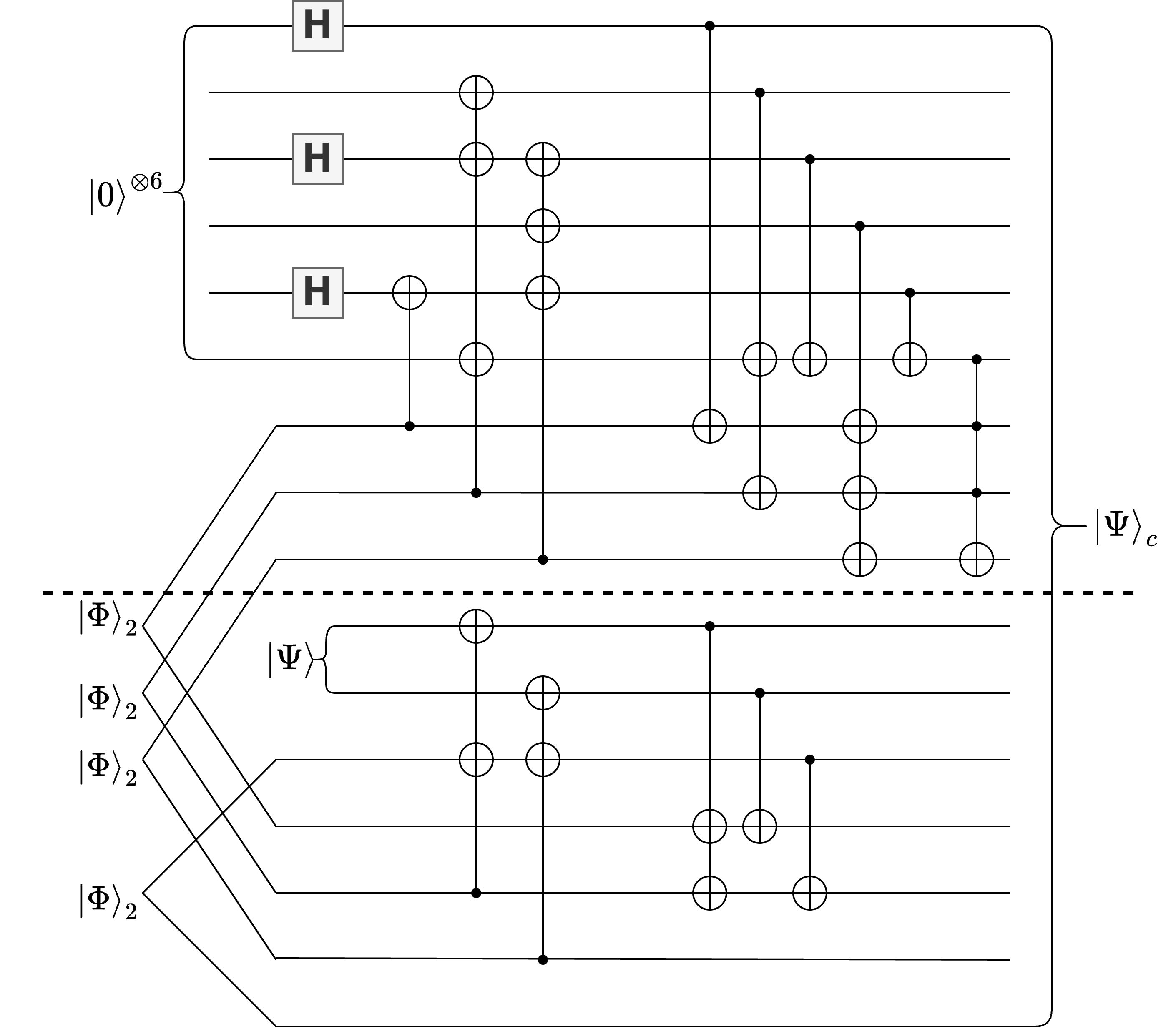}
    \caption{The fault-tolerant encoding circuit of a $[[15,10,2;1]]_2$ entanglement-assisted quantum code designed from the $[[9,4,2;1]]_2$ entanglement assisted quantum code is shown. The information state $\ket{\Psi}$ and the four Bell pairs. The reader must note that corresponding to the three stabilizers, we have three $\ket{\Phi}_2$ preshared entangled states across the blocks of the quantum code that are of the form $\ket{\Phi^{+}}$. The other Bell pair $\ket{\Phi^{+}}$ is shared across the transceiver  towards achieving entanglement assistance. For convenience, we show all these entangled states as $\ket{\Phi}_2$ in the architecture. $\ket{\Psi}$ is encoded using a series of CNOT gates, where three $\ket{\Phi}_2$s are used to make encoder fault-tolerant and the last $\ket{\Phi}_2$ is used for the entanglement assistance of the code. The horizontal dashed line separates qubits into two different blocks of qubit where CNOT gates are applied locally.
    }
    \label{fig:3}
\end{figure*}
where $||$ line separates qubits into two blocks $B_1$ and $B_2$ towards achieving transversality, $|||$ line separates transmitter end and receiver end qubits containing an EPR pair shared between the transmitter ($9^{\mathrm{th}}$ column and last column of $H^{(x)}$ or $H^{(z)}$) and receiver qubits. For the fault-tolerant encoding, we include the extra three EPR pairs in the $\ket{\Phi}_3$ state after the blocks $B_1$ and $B_2$. The parameters of the modified quantum EA code are $Q_F=[[15,10,2;1]]_2$ such that the X and Z stabilizer matrices become
\begin{tiny}
    \begin{align}
     H^{(x)}_F&=\left[\begin{array}{cccccc|ccc||ccc|ccc|||c}
       1 & 0 & 0  & 0 & 0 & 0 & 1&0&0& 0 & 0 & 1 &1& 0&0&1 \\
        0 & 0 & 1  & 0 & 0 & 0 & 0&1&0& 0 & 0 & 1&0&1&0&1\\
        0 & 0 & 0  & 0 & 1 & 0 &  0&0&1 & 0 & 0&1&0&0&1&1\\
        \end{array}\right],\nn\\
      H^{(z)}_F&=\left[\begin{array}{cccccc|ccc||ccc|ccc|||c}
       0 & 1 & 0  & 0 & 0 & 0 & 1&0&0& 0 & 0 & 1 &1& 0&0&1 \\
        0 & 0 & 0  & 1 & 0 & 0 & 0&1&0& 0 & 0 & 1&0&1&0&1\\
        0 & 0 & 0  & 0 & 0 & 1 &  0&0&1 & 0 & 0&1&0&0&1&1\\
        \end{array}\right],
        \end{align}

\end{tiny}
 where $|$ separates entangled qubits and ancille or information qubits in a particular block.
The codeword state $\ket{\psi}$ contains information qubits. We have totally four preshared entangled states (three of them corresponding to the three stabilizers of the code and one of them towards entanglement-assistance). All these preshared quantum states take the form $\ket{\Phi}_2=\ket{\Phi^{+}}$. 

To encode $Q_F$, we first apply $$U_1=\mathrm{CNOT}^{(7,5)}\mathrm{CNOT}^{(8,\{2,3,6\})}\mathrm{CNOT}^{(9,\{3,4,5\})}$$ operator in the first block changing the stabilizer matrices as follows:

  \begin{tiny}
    \begin{align}
     H^{(x)}_F=&\left[\begin{array}{cccccc|ccc||ccc|ccc|||c}
       1 & 0 & 0  & 0 & 1 & 0 & 1&0&0& 0 & 0 & 1 &1& 0&0&1 \\
        0 & 1 & 0  & 0 & 0 & 1 & 0&1&0& 0 & 0 & 1&0&1&0&1\\
        0 & 0 & 1  & 1 & 0 & 0 &  0&0&1 & 0 & 0&1&0&0&1&1\\
        \end{array}\right],\nn\\
      H^{(z)}_F=&\left[\begin{array}{cccccc|ccc||ccc|ccc|||c}
       0 & 1 & 0  & 0 & 0 & 0 & 1&1&0& 0 & 0 & 1 &1& 0&0&1 \\
        0 & 0 & 0  & 1 & 0 & 0 & 0&1&1& 0 & 0 & 1&0&1&0&1\\
        0 & 0 & 0  & 0 & 0 & 1 &  0&1&1 & 0 & 0&1&0&0&1&1\\
        \end{array}\right].
\end{align}
\end{tiny}

Next, we apply $$U_2=\mathrm{CNOT}^{(14,\{10,12\})}\mathrm{CNOT}^{(15,\{11,12\})}$$ 
in the second block such that stabilizer matrices are transformed as follows:

 \begin{tiny}
    \begin{align}
     H^{(x)}_F=\left[\begin{array}{cccccc|ccc||ccc|ccc|||c}
       1 & 0 & 0  & 0 & 1 & 0 & 1&0&0& 0 & 0 & 1 &1& 0&0&1 \\
        0 & 1 & 0  & 0 & 0 & 1 & 0&1&0& 1 & 0 & 0&0&1&0&1\\
        0 & 0 & 1  & 1 & 0 & 0 &  0&0&1 & 0 & 1&0&0&0&1&1\\
        \end{array}\right],\nn\\
      H^{(z)}_F=\left[\begin{array}{cccccc|ccc||ccc|ccc|||c}
       0 & 1 & 0  & 0 & 0 & 0 & 1&1&0& 0 & 0 & 1 &1& 1&1&1 \\
        0 & 0 & 0  & 1 & 0 & 0 & 0&1&1& 0 & 0 & 1&0&0&1&1\\
        0 & 0 & 0  & 0 & 0 & 1 &  0&1&1 & 0 & 0&1&0&1&0&1\\
        \end{array}\right].
\end{align}
\end{tiny}
The rank of the submatrix $W_{B_1}$ containing the $\{7,8,9\}$ columns of $H^{(z)}_F$ is not full rank. So, to obtain the operator of $H^{(z)}_F$, we include the last nonzero column of $H^{(z)}_{F_{B_1}}$, which contains columns of $H^{(z)}_F$ indexed in $B_1$ and solve $\left[H^{(z)}_{F_{*,6}\,W_{B_1}}\right]X^{(B_1)}=H^{(z)}_{F_{*,B_1\backslash\{6\}}}$ 
$$\left[\begin{array}{ccccc}
   0& 1 & 1 & 0  \\
     0& 0 & 1 & 1  \\
       1&0 & 1 & 1 \\
\end{array}\right]X^{(B_1)}=\left[\begin{array}{ccccccc}
     1& 1& 0& 0& 0\\
     0& 1& 0& 0& 0\\
     0& 0& 1& 0& 1
\end{array}\right],$$
$$X^{(B_1)}=\left[\begin{array}{cccccccc}
    0 & 1 & 1 & 0 & 1  \\
    1 & 0 & 0 & 1 & 0  \\
    0 & 1 & 0 & 1 & 0  \\
    0 & 0 & 0 & 1 & 0 
\end{array}\right].$$
Using $X^{(B_1)}$, we apply
\begin{align*}
    O_1=&\mathrm{CNOT}^{(1,7)}\mathrm{CNOT}^{(2,\{6,8\})}\mathrm{CNOT}^{(3,6)}\mathrm{CNOT}^{(4,\{7,8,9\})}\\&\mathrm{CNOT}^{(5,6)}.
\end{align*}
and the stabilizer matrices become
 \begin{tiny}
    \begin{align}
     H^{(x)}_F&=\left[\begin{array}{cccccc|ccc||ccc|ccc|||c}
       1 & 0 & 0  & 0 & 1 & 1 & 0&0&0& 0 & 0 & 1 &1& 0&0&1 \\
        0 & 1 & 0  & 0 & 0 & 0 & 0&0&0& 1 & 0 & 0&0&1&0&1\\
        0 & 0 & 1  & 1 & 0 & 1 &  1&1&0 & 0 & 1&0&0&0&1&1\\
        \end{array}\right],\nn\\
      H^{(z)}_F&=\left[\begin{array}{cccccc|ccc||ccc|ccc|||c}
       1 & 0 & 0  & 0 & 0 & 0 & 1&1&0& 0 & 0 & 1 &1& 1&1&1 \\
        0 & 1 & 0  & 1 & 0 & 0 & 0&1&1& 0 & 0 & 1&0&0&1&1\\
        0 & 0 & 1  & 0 & 1 & 1 &  0&1&1 & 0 & 0&1&0&1&0&1\\
        \end{array}\right].
\end{align}
\end{tiny}
Since $9^{\mathrm{th}}$ column of $H^{(x)}_F$ is zero,  we apply $R^{(x)}_1=\mathrm{CNOT}^{(\{6,7,8\},9)}$ and make it nonzero as follows:
 \begin{tiny}
    \begin{align}
     H^{(x)}_F&=\left[\begin{array}{cccccc|ccc||ccc|ccc|||c}
       1 & 0 & 0  & 0 & 1 & 1 & 0&0&1& 0 & 0 & 1 &1& 0&0&1 \\
        0 & 1 & 0  & 0 & 0 & 0 & 0&0&0& 1 & 0 & 0&0&1&0&1\\
        0 & 0 & 1  & 1 & 0 & 1 &  1&1&1 & 0 & 1&0&0&0&1&1\\
        \end{array}\right],\nn\\
      H^{(z)}_F&=\left[\begin{array}{cccccc|ccc||ccc|ccc|||c}
       1 & 0 & 0  & 0 & 0 & 0 & 1&1&0& 0 & 0 & 1 &1& 1&1&1 \\
        0 & 1 & 0  & 1 & 0 & 1 & 1&0&1& 0 & 0 & 1&0&0&1&1\\
        0 & 0 & 1  & 0 & 1 & 0 &  1&0&1 & 0 & 0&1&0&1&0&1\\
        \end{array}\right].
\end{align}
\end{tiny}
The submatrix $W_{B_2}$ containing $\{13,14,15\}$ columns of $H^{(z)}_F$ is full rank, so we solve the equation $\mathcal{E}_2 : W_{B_2}X^{(B_2)}=H^{(z)}_{F_{*,\Tilde{B}_2}}$, where $\Tilde{B}=\{3+b:b\in B_2\}$. The equation $\mathcal{E}_2$ is solved as follows:
$$\left[\begin{array}{cccc}
    1 & 1 & 1  \\
      0 & 0 & 1  \\
       0 & 1 & 0  \\
\end{array}\right]X^{(B_2)}=\left[\begin{array}{cccc}
     0& 1& 1\\
     0& 0& 0\\
     1& 0& 1
\end{array}\right],$$  to obtain solution $X^{(B_2)}=\left[\begin{array}{cccc}
     1& 1 & 0 \\
     1& 0 & 1 \\
     0& 0 & 0 
\end{array}\right].$
Finally, using $X^{(B_2)}$, we apply $$O_2=\mathrm{CNOT}^{(10,\{13,14\})}\mathrm{CNOT}^{(11,13)}\mathrm{CNOT}^{(12,14)}.$$
The X and Z stabilizer matrices due to $O_2$ change as follows:
 \begin{tiny}
    \begin{align}
     H^{(x)}&=\left[\begin{array}{cccccc|ccc||ccc|ccc|||c}
       1 & 0 & 0  & 0 & 1 & 1 & 0&0&1& 0 & 0 & 1 &1& 1&0&1 \\
        0 & 1 & 0  & 0 & 0 & 0 & 0&0&0& 1 & 0 & 0&1&0&0&1\\
        0 & 0 & 1  & 1 & 0 & 1 &  1&1&1 & 0 & 1&0&1&0&1&1\\
        \end{array}\right],\nn\\
      H^{(z)}&=\left[\begin{array}{cccccc|ccc||ccc|ccc|||c}
       1 & 0 & 0  & 0 & 0 & 0 & 1&1&0& 0 & 1 & 0 &1& 1&1&1 \\
        0 & 1 & 0  & 1 & 0 & 1 & 1&0&1& 0 & 0 & 1&0&0&1&1\\
        0 & 0 & 1  & 0 & 1 & 0 &  1&0&1 & 1 & 0&0&0&1&0&1\\
        \end{array}\right].
\end{align}
\end{tiny}
The encoding operator $\mathcal{E}=O_2R^{(x)}_1O_1U_2U_1$ is shown in Figure \ref{fig:3}.
\subsection{Error Propagation Analysis}
For the fault-tolerant encoder, we derive a bound on the probability of error propagation in the following lemma.
\begin{lemma} \label{lemma:6} For a $[[n+g\rho,k+g\rho,d]]_2$ CSS code with X and Z stabilizer matrices $H^{(x)}_F$ and $H^{(z)}_F$, the probability of error propagation due to CNOT gates for a fault-tolerant encoder is upper bounded by $P_{F} \leq 1 -P_{F_1} P_{F_2}$, where
\begin{scriptsize}
\begin{align}
    P_{F_1} &\geq \prod_{i=1}^{\rho_1}\prod_{j=1}^{g} \left(1-\frac{2p}{3}+ \sum_{c \in M^{(x)}_{i,j}} {w^{(x)}_{i,j} \choose c}\left(\frac{2p}{3}\right)^c \nn \right.  \left.\left(1-\frac{2p}{3}\right)^{w^{(x)}_{i,j}-c}\right),\nn\\
    P_{F_2} &\geq \prod_{i=1}^{\rho_2}\prod_{j=1}^{g} \left(1-\frac{2p}{3}+ \sum_{c \in M^{(z)}_{i,j}} {w^{(z)}_{i,j} \choose c}\left(\frac{2p}{3}\right)^c \nn \right.  \left.\left(1-\frac{2p}{3}\right)^{w^{(z)}_{i,j}-c}\right),
\end{align}
\end{scriptsize}
$p$ is the probability of depolarizing noise channel $\pi(p)$, $w^{(x)}_{i,j}$ $\left(\mathrm{or}\, w^{(z)}_{i,j}\right)$ is the Hamming weight of $H^{(x)}_{i,B_j}+M_{i,B_j}$ $\left(\mathrm{or}\, H^{(z)}_{i,B_j}+N_{i,B_j}\right)$ and $M^{(x)}_{i,j}=\{0,2,\cdots ,w^{(x)}_{i,j}\}\, \left( \mathrm{or}\,M^{(z)}_{i,j}=\{0,2,\cdots ,w^{(z)}_{i,j}\}\right)$. The matrices $M_{i,B_j}$ and $N_{i,B_j}$ are the submatrices of $H^{(x)}$ and $H^{(z)}$ before encoding containing the columns from $B_j$\footnote{For $[[15,10,2;1]]_2$ code, $M_{B_1}$ and $N_{B_1}$ contains the first six columns of $H^{(x)}$ and $H^{(z)}$ in equation (\ref{Eq:Befor_enc_mat}).}.
\end{lemma}
\begin{IEEEproof}
We prove the lemma in the Appendix of arxiv version of the paper. \\
\end{IEEEproof}
Using Lemma \ref{lemma:5} and \ref{lemma:6}, we can evaluate the minimum number blocks required to outperform the non-fault-tolerant encoder using fault-tolerant encoder in terms of error propagation, i.e., $P_{F}<P_{NF}$.

\section{Conclusions}\label{Sec.5}
We proposed non-fault-tolerant and fault-tolerant encoders for quantum LDPC encoders. Dual-containing LDPC codes from a good family of classical LDPC codes can be extended using error-free EPR pairs to avoid short cycles. The entanglement-assisted extended LDPC code can be divided into block of qubits that share  common entanglement through a multipartite state. Using preshared multipartite entanglement across the stabilizer generators of the code, one can achieve transversal implementation. The fault-tolerant design is also proven efficient in terms improved error propagation probability over the non-fault-tolerant design, useful towards practice. 
\section*{Acknowledgment} This work is financially supported by the Department of Science and Technology (DST), Govt. of India, 
 with Ref No. SERB/F/3132/2023-2024.
\clearpage
\bibliography{funbibfile}
\clearpage
\appendix
In this Appendix, we provide proofs of the results that
appear in the main body of the paper. We also provide an example for the fault-tolerant encoder. 
\subsection{Proofs of the results in the main text}\label{Appendix:1}

First, we prove that in a non-fault tolerant encoder provided in Section \ref{Section:fault} the operator $U$ will make first $\rho_1$ columns of the $H^{(z)}$ matrix zero. Any gate applied on a codeword will change the complete set of stabilizer generators and $U$ has $\rho_1$ number of CNOTs that are controlled from the first $\rho_1$ qubits. We can thus transform the first $\rho_1$ columns of $H^{(z)}$ to zero. The effect on $H^{(z)}$ is due to the property of CSS codes. If $\rho_1\leq\rho_2$, then
\begin{align*}
\Tilde{H}^{(x)}_{i,*}\Tilde{H}^{(z)^{\mathrm{T}}}_{j,*}&=I^{(x)}_{i,*}I^{(z)^{\mathrm{T}}}_{j,*}+A^{(x)}_{i,*}A^{(z)^{\mathrm{T}}}_{j,*}+B^{(x)}_{i,*}B^{(z)^{\mathrm{T}}}_{j,*},\\ &=0 \mod (2),
\end{align*}
for $1\leq i,j\leq\rho_1$. The symbol $\mathrm{T}$ stands for the transpose of a matrix. This implies $$A^{(x)}_{i,*}A^{(z)^{\mathrm{T}}}_{j,*}+B^{(x)}_{i,*}B^{(z)^{\mathrm{T}}}_{j,*}=\delta_{j,i} \mod (2).$$
We can further simplify it as 
\begin{equation}
    \sum_{k:A^{(x)}_{i,k}=1} A^{(z)}_{j,k}+\sum_{k:B^{(x)}_{i,k}=1} B^{(z)}_{j,k}=\delta_{j,i} \mod (2).
    \label{eq:4}
\end{equation}
Since the target columns are $C_i=\{j: A^{(x)}_{i,j}=1\, \mathrm{or}\, B^{(x)}_{i,j}=1\}$ corresponding to the $U_i$ gate, using equation (\ref{eq:4}), we get sum of all the columns of $\Tilde{H}^{(z)}$ indexed in $C_i$ as $\ubar{s}_{i}=[\delta_{j,i}]_{j=1}^{\rho_2}$ present at the target positions. Since the Pauli Z operator on the target qubits will be transferred to the control qubit, the $\ubar{s}_i$ column will be added to the control column of $I^{(z)}_{*,i}$ (both being ones at location $i$), thereby, nulling it. Similarly,  If $\rho_2<\rho_1$, then equation (\ref{eq:4}) remains the same for $1\leq i,j\leq \rho_2$, but for $\rho_1 \geq i> \rho_2$ and $j\leq \rho_2$ equation (\ref{eq:4}) becomes 
\begin{equation}
    \sum_{k:A^{(x)}_{i,k}=1} A^{(z)}_{j,k}+\sum_{k:B^{(x)}_{i,k}=1} B^{(z)}_{j,k}=A^{(z)}_{j,i} \mod (2).
    \label{eq:5}
\end{equation}
Therefore, using the same argument given for $\rho_1\leq\rho_2$ case, the first $\rho_1$ columns of $H^{(z)}$ become zero from operator $U$.
\\
\begin{IEEEproof}[Proof of the Lemma \ref{lemma:5}]
    Consider a quantum CSS code $Q=[[n,k,d]]_2$ with X and Z stabilizer matrices $\Tilde{H}^{(x)}=\left[I^{(x)}|A^{(x)}|B^{(x)}\right]$ and $\Tilde{H}^{(z)}=\left[I^{(z)}|A^{(z)}|B^{(z)}\right]$ such that $H^{(x)}H^{(z)^{\mathrm{T}}}=\textbf{0}\mod (2)$. 
    
    Let the control and target qubits be affected by the depolarizing noise channel $\pi(p)$. First, we iterate over the rows of $H^{(x)}$ and apply $U_i=\mathrm{CNOT}^{(i,c_i)}$ gate, for the $i^{\mathrm{th}}$ iteration, having the columns of $I^{(x)}$ as control and the nonzero columns of $\left[A^{(x)}|B^{(x)}\right]$ as targets. Let $p^{(AB)}_i$ be the probability such that $U_i$ causes no error propagation. Consider that the control qubit of the CNOT in $U_i$ is affected by either X, Y, or Z errors with probability (w.p.) $\frac{p}{3}$ and I w.p. $1-p$. As I and Z acting on the control qubit w.p. $1-\frac{2p}{3}$ do not propagate to targets, no error propagation happens. However, X or Y errors on the control qubit w.p. $\frac{2p}{3}$ will generate X errors on target qubits as well, resulting in a propagation of error. 
    
    Now, if the target qubits of CNOT in $U_i$ are affected by X, Y or Z, then I or X w.p. $1-\frac{2p}{3}$ causes no propagation of errors to the control qubit, but when an odd number of target qubits are affected with Y or Z, w.p. $\frac{2p}{3}$, they generate a Z error on the control qubit. Therefore, using the binomial distribution, we get 
    \begin{equation}
        p^{(AB)}_i = 1-\frac{2p}{3} + \sum_{c \in M^{(AB)}_i} {w^{(AB)}_i \choose c} \left(\frac{2p}{3}\right)^c \left(1-\frac{2p}{3}\right)^{w^{(AB)}_i-c},
    \end{equation}
  $w^{(AB)}_i$ is the Hamming weight of the row $\left[A^{(x)}_{i,*}|B^{(x)}_{i,*}\right]$ and $M^{(AB)}_i=\{0,2,\cdots ,w^{(AB)}_i\}$, i.e., even terms upto $w^{(AB)}_i$. Thus, we get the total probability of no error propagation due to operator $U=\prod_{i=1}^{\rho_1} U_i$ as follows:
    \begin{align}
        P_{NF_1} \geq \prod_{i=1}^{\rho_1}& \left(1-\frac{2p}{3} + \sum_{c \in M^{(AB)}_i} {w^{(AB)}_i \choose c} \left(\frac{2p}{3}\right)^c \nn \right. \\& \left.\left(1-\frac{2p}{3}\right)^{w^{(AB)}_i-c}\right).
        \label{eq:7}
    \end{align}
    The lower bound on $P_{NF_1}$ is because we excluded the cases of self-correction of errors. For example, if the first and second qubit in a three-qubit case are affected with X errors, then the propagated error on the third qubit due to $\mathrm{CNOT}^{(1,3)}$ is nullified by the propagation of error due to $\mathrm{CNOT}^{(2,3)}$, i.e., 
    \begin{align*}
        \mathrm{CNOT}^{(2,3)}\mathrm{CNOT}^{(1,3)}(X\otimes X \otimes I)\mathrm{CNOT}^{(1,3)}\mathrm{CNOT}^{(2,3)}\nn\\=\mathrm{CNOT}^{(2,3)}(X\otimes X \otimes X)\mathrm{CNOT}^{(2,3)}=X\otimes X\otimes I.
    \end{align*}
    Similarly, we can calculate the probability of 
 no error propagation due to CNOTs in operator $W$ applied to make $B^{(z)}$ submatrix zero, as follows:
    \begin{align}
        P_{NF_2} \geq &\prod_{i=1}^{w} \left(1-\frac{2p}{3} + \sum_{c \in M^{(A)}_i} {w^{(A)}_i \choose c} \left(1-\frac{2p}{3}\right)^c \nn \right. \\& \left.\left(\frac{2p}{3}\right)^{w^{(A)}_i-c}\right),
        \label{eq:8}
    \end{align}
    where $w=n-\rho_1-\rho_2$, $w^{(A)}_i$ is the Hamming weight of $A^{(z)}_{i,*}$ and $M^{(A)}_i=\{0,2,\cdots ,w^{(A)}_i\}$.
    Finally, the probability of error propagation in a non-fault-tolerant encoder is $P_{NF} \leq 1- P_{NF_1} P_{NF_2}$.
\end{IEEEproof}
\begin{IEEEproof}[Proof of Lemma \ref{lemma:6}]
We follow the steps similar to Lemma~\ref{lemma:5}. Consider the $[[n+g\rho,k+g\rho,d]]_2$ quantum code. With depolarizing noise probability $p$, every $\mathrm{CNOT}^{(c^{(x)}_{i,j},t^{(x)}_{i,j})}$ gate in operator $U_1$ applied over $H^{(x)}_{F_{i,b_j}}$  ( refer to Section \ref{Section:fault}) has a probability of no error propagation $p^{(x)}_{i,j}$ given by
\begin{equation}
    p^{(x)}_{i,j} =1-\frac{2p}{3}+ \sum_{c \in M^{(x)}_{i,j}} {w^{(x)}_{i,j} \choose c}\left(\frac{2p}{3}\right)^c \left(1-\frac{2p}{3}\right)^{w^{(x)}_{i,j}-c},
\end{equation}   
where $M^{(x)}_{i,j}=\{0,2,\cdots,w^{(x)}_{i,j}\}$ and $w^{(x)}_{i,j}$ is the Hamming weight $K_{i,b_j}=H^{(x)}_{i,b_j}+M_{i,b_j}$, i.e., the weight of the $i^{\mathrm{th}}$ row in the $b_j^{\mathrm{th}}$ block for the $K_{i,b_j}$ matrix.
The total probability of no error propagation due to operator $U$ is given by
\begin{equation}
    P_{F_1} \geq \prod_{i=1}^{\rho_1}\prod_{j=1}^{g} p^{(x)}_{i,j}.
\end{equation}
For the $\mathrm{CNOT}^{(c^{(z)}_{i,j},t^{(z)}_{i,j})}$ in operator $O$ in Section \ref{Section:fault}, the probability of no error propagation is given by
\begin{equation}
    p^{(z)}_{i,j} = 1-\frac{2p}{3}+ \sum_{c \in M^{(z)}_{i,j}} {w^{(z)}_{i,j} \choose c}\left(\frac{2p}{3}\right)^c \left(1-\frac{2p}{3}\right)^{w^{(z)}_{i,j}-c},
\end{equation}   
where $M^{(z)}_{i,j}=\{0,2,\cdots,w^{(z)}_{i,j}\}$ and $w^{(z)}_{i,j}$ is the Hamming weight $H^{(z)}_{i,b_j}+N_{i,b_j}$.
The total probability of no error propagation due to operator $O$ is given by
\begin{equation}
    P_{F_2} \geq \prod_{i=1}^{\rho_2}\prod_{j=1}^{g} p^{(z)}_{i,j}.
\end{equation}
Finally, the total probability of error propagation is bounded by $P_F \leq 1-P_{F_1} P_{F_2}$. 
\end{IEEEproof}

\end{document}